# Preparation of the oxypnictides and studies on their superconductivity

Ayaka KAWABATA, Sang Chul LEE, Taketo MOYOSHI, Yoshiaki KOBAYASHI, and Masatoshi SATO*

*Department of Physics, Division of Material Science, Nagoya University, Furo-cho, Chikusa-ku, Nagoya 464-8602 Japan*

$LaFe_{1-y}Co_yAsO_{1-x}F_x$ ($x$=0.11) with various $y$ values were prepared and their electrical resistivities, superconducting diamagnetisms and Hall coefficients have been measured. $^{75}$As- and $^{135}$La-NMR studies have also been carried out In spite of the successful Co-doping, we have not found any meaningful correlation of $T_c$ with $y$, which indicates that the $T_c$-suppression by Co-doping is considered not to be so significant as expected for superconductors with nodes. Even for superconductors without nodes, it may not be easy to expect this small effect on $T_c$, if there are two different (disconnected) Fermi surfaces whose order parameters have opposite signs. The data of the NMR Knight shift indicate that the Cooper pairs are in the singlet state and the spin susceptibility is almost fully suppressed at low temperature.

## 1. Introduction

Superconductivity in $LaFeAsO_{1-x}F_x$ found by Kamihara *et al.* [1] presented a remarkable example of 3*d* electron superconductors and a variety of superconducting compounds having FeAs layers of edge-sharing $FeAs_4$ tetrahedra with the transition temperature $T_c$ higher than 50 K have been found.[2] Because the superconductivity primarily occurs in the 3*d*-electrons of the FeAs layers,[3,4] the magnetic interaction cannot be ignored in the study of the origin of the superconductivity.

We have prepared $LaFe_{1-y}Co_yAsO_{1-x}F_x$ system ($x$=0.11)[5] and measured their electrical resistivities ρ, magnetizations $M_s$ due to the superconducting diamagnetism $M_s$ and Hall coefficients $R_H$. We have also carried out $^{75}$As- and $^{135}$La-NMR measurements.

Here, we present results of the above measurements and argue the symmetry/origin of the superconductivity and the electronic state.[5] The $T_c$-suppression by the Co-impurities is rather weak as compared with that observed for Cu oxides.[6] We also show that the spin component of the Knight shift of $LaFeAsO_{1-x}F_x$ ($x$=0.11) is almost completely suppressed by the superconductivity, confirming our previous data that the system has singlet Cooper pairs.[5]

## 2. Experiments

Polycrystalline samples of $LaFe_{1-y}Co_yAsO_{1-x}F_x$ ($x$=0.11; $0 \le y \le 0.3$) were prepared as described in ref. 5 from initial mixtures of La, $La_2O_3$, $LaF_3$ and FeAs with the nominal molar ratios. A SQUID magnetometer was used to measure $M_s$ and ρ was measured by the four terminal method and their data are in ref. 5. From the $M_s$-$T$ curves we determined the $T_c$ values. $R_H$ was measured for samples with $y \le 0.03$ with the magnetic field $H$= 7 T by rotating samples. The $^{75}$As- and $^{135}$La-NMR measurements were carried out by the standard coherent pulse method, where the nuclear spin-echo intensity $I$ was recorded with the NMR frequency or applied magnetic field changed stepwise.

## 3. Results and Discussion

In the inset of Fig. 1, the lattice parameter $c$ of $LaFe_{1-y}Co_yAsO_{1-x}F_x$ ($x$=0.11) is plotted against $y$. The linear relationship between $c$ and $y$ guarantees that the Co doping

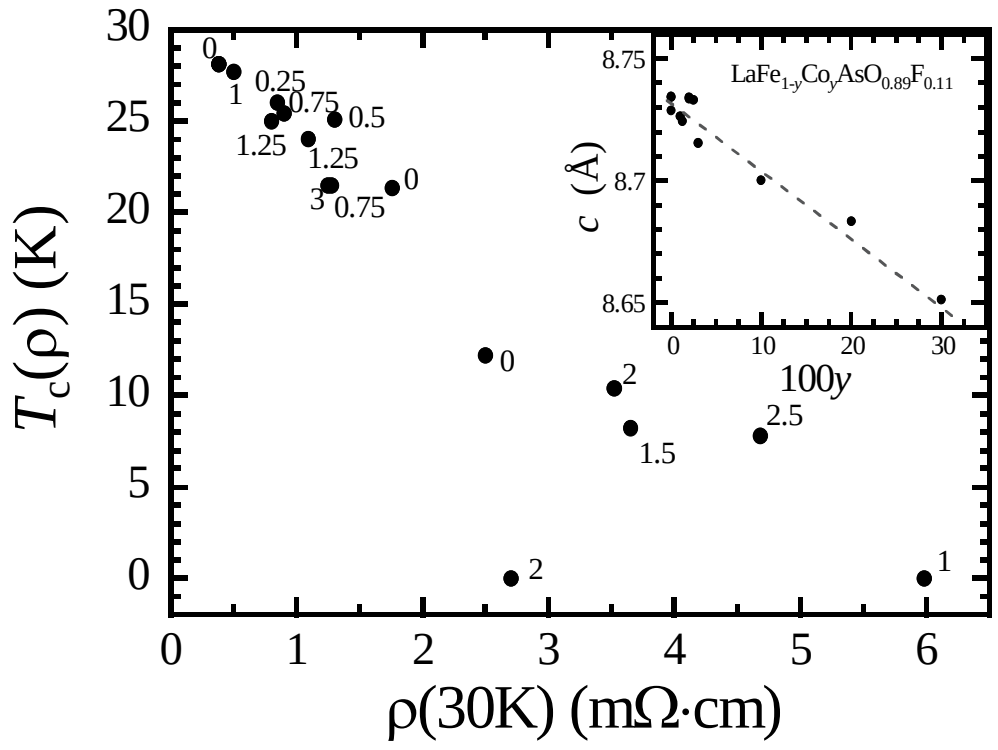

Fig. 1. $T_c$ is plotted against ρ(30 K), where 100$y$ values are attached to the corresponding data. Inset shows the lattice parameter $c$ against 100$y$.

was successfully carried out. In the main panel of Fig.1, $T_c$ is plotted as a function of the resistivity at 30 K, ρ(30 K). While we can find the systematic $T_c$-decrease with increasing ρ(30 K), we do not find any correlation of $T_c$ with $y$, indicating that the $T_c$ value is not primarily determined by the Co concentration $y$ but by the electrical resistivity ρ,[5] and therefore that $T_c$-suppression by impurity scattering is much smaller than expected for order parameters Δ with nodes on the Fermi surface. Even for superconductors without nodes, this small $T_c$-suppression

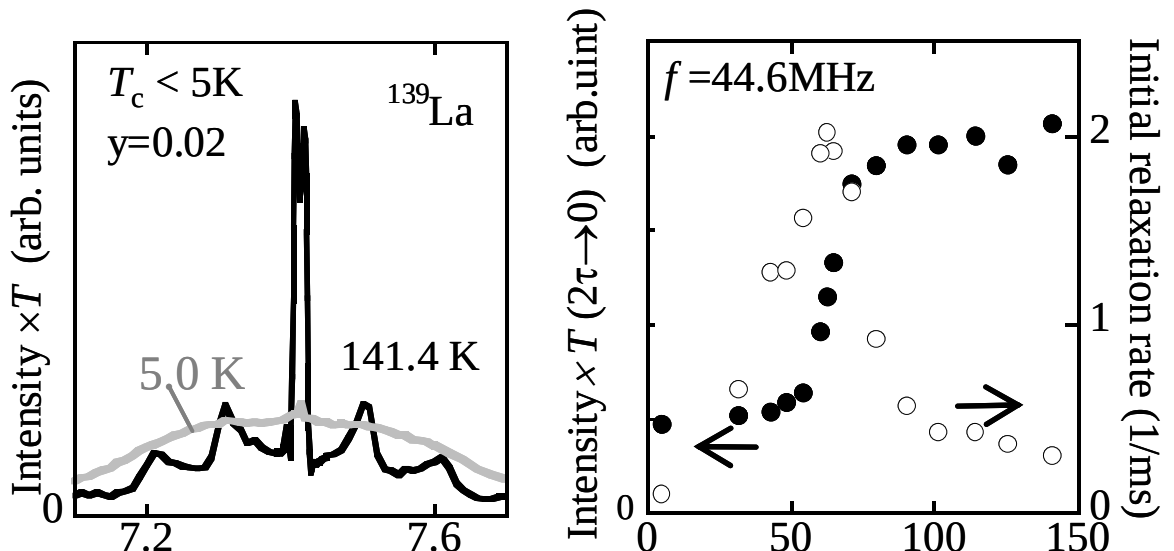

Fig. 2. (left) NMR intensities of $^{139}$La multiplied by $T$ taken at a frequency $f$ = 44.6 MHz for the sample $y$~0.02 ($T_c$<5 K) are shown at two temperatures. (right) $T$ dependences of the NMR intensities multiplied by $T$ (•) at $H$~7.4 T and the initial spin-spin relaxation rate $1/\tau_0$ (°). Effects of magnetic ordering can be seen at ~63 K.

*corresponding author: e43247a@nucc.cc.nagoya-u.ac.jp

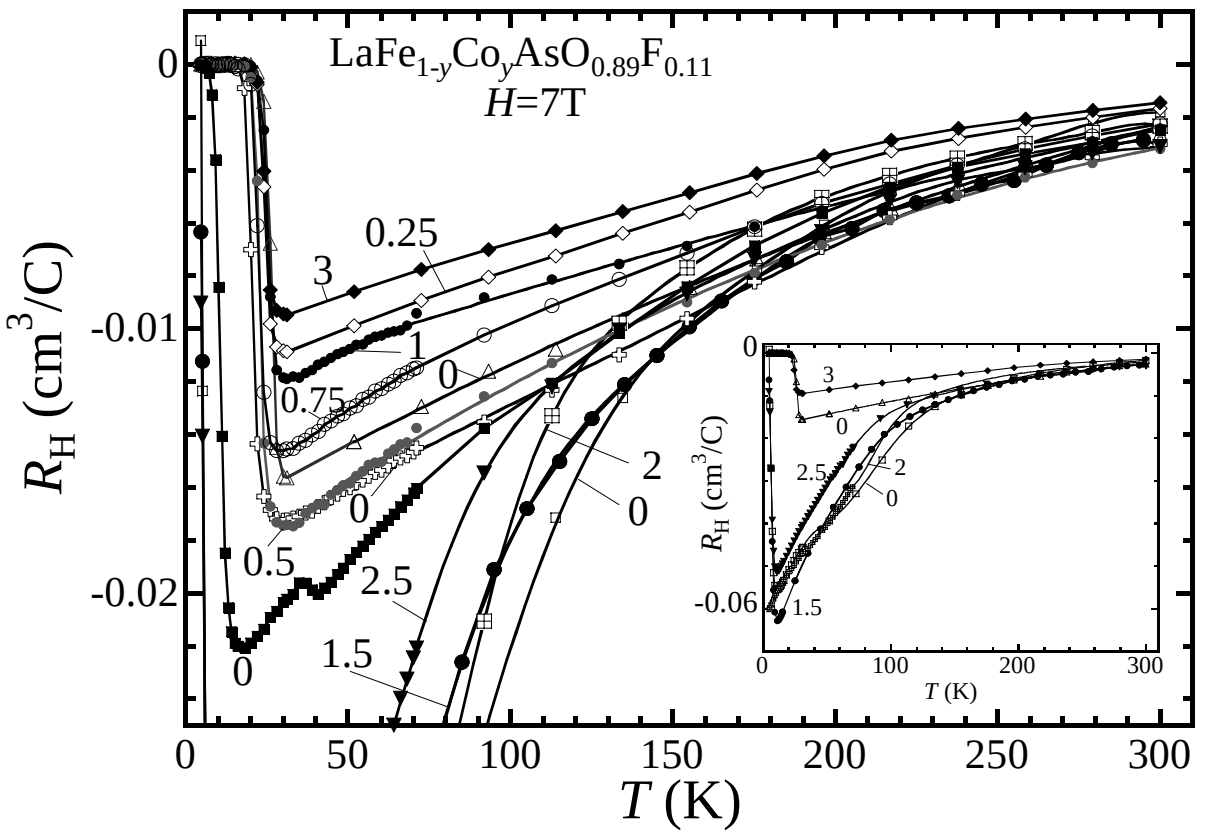

Fig. 3. $T$ dependences of $R_H$ of $LaFe_{1-y}Co_yAsO_{1-x}F_x$ ($x$=0.11) are shown, where 100y values are attached to the corresponding data.

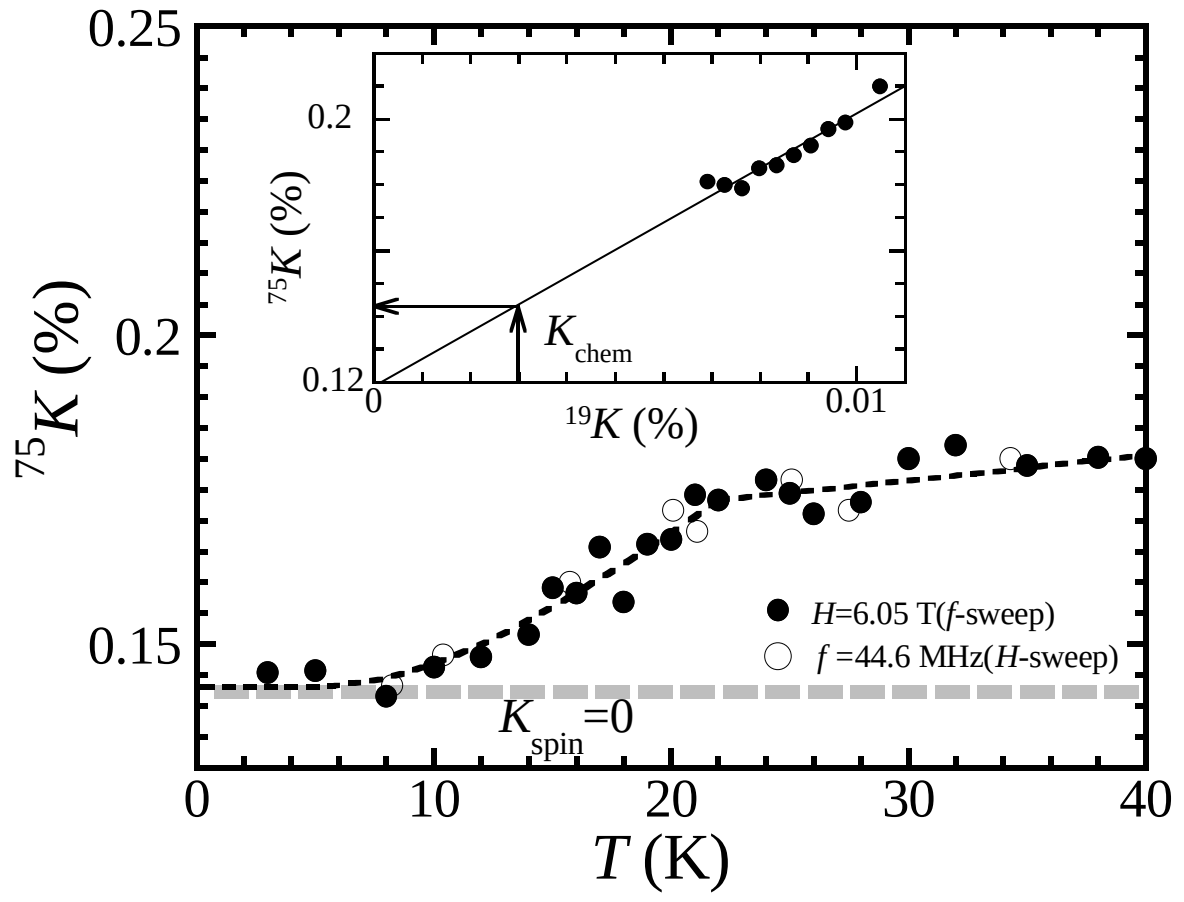

Fig. 4. Data of $^{75}K$ obtained with two different conditions are shown against $T$. The grey broken line corresponds to the spin contribution $^{75}K_{spin}$ =0. Note that the chemical shift of $^{19}K$=0.03%.[12)]

rate may not be easily expected, if there are two different (disconnected) Fermi surfaces whose order parameters have opposite signs. Therefore, this smallness of the $T_c$ suppression rate is important for arguments on the relative signs of Δ on two kinds of Fermi surfaces,[7,8)] and consistent with the superconductivity found in heavily Co-doped systems.[9)] As one of possible origins of the observed weak $T_c$-decrease with ρ(30 K), the loss of the itinerant nature can be considered, though the intrinsic resistivity cannot be accurately estimated because of the grain-boundary effect (detailed discussion are in ref. 5.) With increasing the resistivity, the behavior of the NMR spin-spin relaxation rate $1/\tau$ changes,[5)] and the initial relaxation rate $1/\tau_0$ increases, indicating that the spin fluctuation of the system increases. In Fig. 2, the $T$ dependences of $1/\tau_0$ and spectral intensity multiplied by $T$ (right panel) and the spectral broadening at $T< T_N$ (left panel) are shown for the sample with 100$y$=2.0 ($T_c$<5 K). Evidence for the antiferromagnetic (or SDW) transition at $T_N$~63 K found in the figure, is a further support of this idea.

Figure 3 shows the $T$ dependence of $R_H$ of the $LaFe_{1-y}Co_yAsO_{1-x}F_x$ ($x$=0.11) samples. While $R_H$ does not exhibit any correlation with $y$, the behavior changes systematically with $T_c$, which resembles to that of high $T_c$ Cu Oxides.[10-12)] If the magnetic fluctuation is relevant to the observed strong $T$ dependence of $R_H$ as in Cu oxides, the fact that the characteristic temperature of the $R_H$-$T$ curves is smaller by a factor of 2 than that of Cu oxides is important for the consideration of $T_c$ value.

Figure 4 shows NMR Knight shift of $^{75}$As reported in our previous paper[5)] Here, combining our new data taken up to ~250K for $^{75}$As and those of $^{19}$F,[13)] we estimate the chemical shift of $^{75}$As by the method shown in the inset and indicate it by the grey broken line. The results show that the contribution to the shift $^{75}K_{spin}$ vanishes at low temperature in this system with singlet pairing. We also note that the $T$ dependence of $^{75}K$ does not exhibit the two gap nature and can roughly be explained by the simple Yosida function. It is different from the results of ref.14 for $PrFeAsO_{1-x}F_x$. We think that this discrepancy arises from that in $PrFeAsO_{1-x}F_x$, a possible influence of the Pr moments on the internal field becomes serious.

## 4. Summary

We have shown that the effect of Co impurities on $T_c$ is weak, which restricts the symmetry and the possible origin of the superconductivity. The observed Knight shift indicates the singlet pairing and it roughly obeys the Yosida function.